\documentstyle{article}
\begin{document}

\title{
Bifurcations, Chaos, Controlling and Synchronization of Certain
Nonlinear Oscillators
}
\author{ 
M. Lakshmanan\\
Centre for Nonlinear Dynamics\\
Department of Physics\\
Bharathidasan University\\
Tiruchirapalli - 620 024 \\
India\\
{\it e.mail}: lakshman@kaveri.bdu.ernet.in}
\date{}
\maketitle
\begin{abstract}
In this set of lectures, we review briefly some of the recent
developments in the study of the chaotic dynamics of nonlinear
oscillators, particularly of damped and driven type.  By taking a
representative set of examples such as the Duffing, Bonhoeffer-van
der Pol and MLC circuit oscillators, we briefly explain the various
bifurcations and chaos phenomena associated with these systems.  We
use numerical and analytical as well as analogue simulation methods to study
these systems.  Then we point out how controlling of chaotic motions can
be effected by algorithmic procedures requiring minimal perturbations.
Finally we briefly discuss how synchronization of identically evolving
chaotic systems can be achieved and how they can be used in secure
communications.
\end{abstract}
\newpage
\section{Introduction}
Dynamical systems are often modelled by differential equations when the
flow is continuous.  For an N-particle system moving under the influence
of (constraint free) internal and external forces, $\vec F_i = 1,2, ...N$,
Newton's equations of motion, 
$$m_i \frac{d^2 \vec r_i} {dt^2} = \vec F_i \left (\vec r_1,\vec r_2, 
\ldots \vec r_N, \frac{d\vec r_1} {dt}, \ldots  \frac{d\vec r_N}{dt}, 
t \right), \,\, i=1,2,\ldots N, \eqno (1) $$
where $\vec r_i \equiv \vec r_i(t)$ is the position vector of the $i$th particle
in an inertial frame, subjected to $6N$ initial conditions $\vec r_i(0)$ and
$\left (\frac{d\vec r_i} {dt}\right )_{t=0}$, constitute a deterministic  
description of the system.  As long as the initial position and velocities
as well as the nature of the forces are precisely prescribed, by solving 
the system of $6N$ coupled second order ordinary differential equations 
(ODEs) the future evolution can be predicted with as much accuracy as
desired.  Of course the Laplace dictum of complete determinism fails
when statistical fluctuations and quantum effects are present.  Barring the 
presence of these forces, one might expect that the solution of the 
initial value problem of (1) should lead to complete predictability.
However, in recent times it has been clearly demonstrated that a new
kind of indeterminism or unpredictability can arise when the forces acting
on the system are nonlinear (even when statistical and quantum forces
are absent) in suitable form.

When the forces $\vec F_i$ in (1) are linear, we only have to solve a
system of linear differential equations
wherein the superposition principle holds good.  Any small deviation
in the initial conditions or numerical errors will then grow at the 
worst only linearly, so the dynamics of the system can be exactly
predicted. However when the forces $\vec F_i$ are nonlinear, one has to solve
nonlinear differential equations.  Then for suitable nonlinearities, one 
typically finds the possibility that any small deviation in the 
prescription of initial conditions or round-off error at any stage 
of the calculation can multiply quickly, leading to an exponential divergence
of trajectories corresponding to nearby initial conditions.  Or, in other
words, the system can be extremely sensitively dependent on initial 
conditions leading to complex or chaotic motions.  Physically this 
means that any small fluctuation during the evolution can lead to
realizable macroscopic effects, the so-called `butterfly effect' [1,2].
Mathematically, the phase trajectory might end up asymptotically in a
strange attractor of zero volume but of fractional dimensionality in the
case of dissipative systems, or it can fill the whole phase space 
densely in the case of conservative systems.

The notion of complexity and chaos can be realized even in low-dimensional
nonlinear systems.  Particularly, damped and driven oscillators are
physically realizable examples exhibiting the immense varieties of
bifurcations and chaos.  They are often modelled by a single, second-order,
nonlinear differential equation (or equivalent first order system)
with periodic inhomogeneities.  These
include the Duffing oscillator, the Bonhoeffer-van der Pol oscillator,
the Duffing-van der Pol oscillator, the Murali-Lakshmanan-Chua electronic
circuit, the damped driven pendulum, the driven Morse oscillator and so 
on [3,4].  In this set of lectures, we will consider the extremely varied
bifurcations exhibited by these systems and the various types of
regular and chaotic motions admitted by them and some of their applications.

\section {Linear and nonlinear oscillators}
There are obvious characteristic differences between the dynamics of 
linear and nonlinear oscillators.  We will briefly consider these
aspects with simple examples.

\subsection {Linear oscillators}
We will point out how predictability becomes an inherent property of linear 
oscillator systems.  We will also consider the type of attractors exhibited 
by these systems in this section.

\subsubsection {Free linear harmonic oscillator}
The ubiquitous linear harmonic oscillator of unit mass satisfying the 
equation of motion
$$ \frac{d^2x}{dt^2}+\omega_0^2 x=0, \,\,\, \omega _0^2 >0 \eqno(2)$$
and the initial conditions $ x(0) = A$, $  (dx/dt)_{t=0} = 0 $ 
admits the solution
$$ x(t) = A \cos \omega _0 t \eqno(3) $$
such that
$$ E = \frac{1}{2} \dot x^2 + \frac{1}{2}\omega _0^2 x^2 = \frac{1}{2} 
\omega_0^2 A^2 \,\,\,\, \, \,(. = \frac{d}{dt}) \eqno(4) $$
is a constant.  The corresponding phase trajectories in the phase space
($(x$-$\dot x)$-space) are concentric ellipses with the origin $(0,0)$ being
a centre type (elliptic) fixed point (Fig. 1a).
\subsubsection {Damped linear harmonic oscillator}
When viscous damping force is present, the harmonic oscillator equation
becomes
$$ \frac{d^2x}{dt^2} +\alpha \frac{dx}{dt} + \omega _0^2 x = 0, \eqno (5)$$ 
where $\alpha$ $(>0)$ is the damping parameter.  For the same initial 
conditions, $x(0)=A$, $\dot x(0)= 0$, the solution for the physically
interesting underdamped case $(\alpha <2 \omega_0)$ becomes
$$ x(t) = A\frac{\sqrt{\alpha^2/4+C^2}}{C} \exp{\left (-\frac{\alpha t}{2}
\right )}\cos (Ct-\delta), \eqno(5a) $$
where 
$$ C = \sqrt {\omega_0^2 - \frac{\alpha^2}{4}}, \,\,\, \delta = 
\tan ^{-1}\frac{\alpha}{2C}. \eqno(5b) $$
Solution (5) represents a damped oscillatory motion and energy is
dissipated.  The corresponding phase trajectory is a spiral asymptotically
approaching the fixed point at the origin which is now a stable focus (Fig. 1b).
\subsubsection {Damped and driven linear oscillator}
When an external sinusoidal force is added to the system, we have
$$ \frac{d^2x}{dt^2} + \alpha \frac{dx}{dt} + \omega_0^2x = f \sin\omega t. 
\eqno(6)$$
Again for the same set of initial conditions as above, we have the solution
$$ x(t) = A_t \frac {\sqrt{\alpha ^2 /4+C^2}}{C} \exp\left (\frac{-\alpha t}
{2}\right )\cos(Ct-\delta)+A_p \cos(\omega t-\gamma), \eqno(7a)$$
where
$$A_p = \frac{f}{\sqrt{(\omega_0 ^2 - \omega ^2)^2+\alpha ^2 \omega ^2}}, \,\,\, 
\gamma = \tan^{-1}\left( \frac{\alpha \omega}{(\omega_0 ^2 - \omega ^2)}\right),
\eqno(7b)$$
and $A_t$ is chosen so as to satisfy the initial condition $x(0) = A$, 
$\dot {x}(0) = 0$.  The first term in (7) is obviously a transient which 
decays for large $t$.  Thus asymptotically the system oscillates with a
frequency equal to that of the external periodic force.  At the resonance value 
$\omega  = \omega_0$, the amplitude of oscillation $A_p$ takes the 
maximum value.  The associated phase trajectory corresponds asymptotically
to an attractor which is now a limit cycle as depicted in Fig. 1(c).
In all the above cases, the initial value problem is exactly solvable, the 
motion is completely regular and the future is fully predictable, given the
initial state.  Trajectories corresponding to nearby initial conditions
continue to be nearby, leading to complete predictability.  The attractors
are only fixed points or closed curves corresponding to regular motions.
\subsection {Nonlinear oscillators}
As a prelude to understanding the typical behaviour of damped and driven
nonlinear oscillators, we ask how the dynamics of the 
previously studied linear oscillators are modified when a nonlinear
spring force of cubic form is added so that the equation of motion becomes
$$ \frac {d^2x}{dt^2}+\alpha \frac{dx}{dt}+\omega_0^2x+\beta x^3 = 
f\sin\omega t. \eqno(8) $$ 
\subsubsection {Forced oscillations - Resonances $(f\neq 0)$}
One immediately observes that the behaviour of nonlinear oscillators 
of the form (8) can be qualitatively quite different from that of the
linear case, even for very small $\beta$.  Near the primary resonance
$\omega \approx \omega_0$, $\beta\ll 1$, one can write the lowest order 
approximate solution in the form
$$x(t)=A\cos(\omega t+\delta), \eqno(9) $$
where $ \delta $ is a constant to be fixed.  Using (9) in (8) one 
obtains the resonance curve 
$$ \left [(\omega_0^2 - \omega^2)A + \frac{\beta}{4} \beta A^3\right ]^2 + 
(\alpha A \omega)^2 = f^2. \eqno(10) $$
A typical form of this curve is shown in Fig. 2 for $\beta >0$, which clearly 
shows the associated hysteresis phenomenon.  In addition one also 
observes secondary resonances (subharmonic and superharmonic) at
frequencies $ \omega \approx n\omega_0$, and $ \omega = 
\frac{\omega_0}{n}$, $n=1,2,3,\ldots$
\subsubsection {Nonlinear oscillations and bifurcations [1-7]}
However, to obtain a complete picture of the dynamics, one has to have
recourse to detailed numerical analysis, using for example the fourth
order Runge-Kutta integration method.  As a result one obtains a 
rich variety of bifurcation phenomena, namely successive qualitative 
(and quantitative) changes in the nature of oscillations as the 
nonlinearity parameter (or another control parameter say $f$ or $\omega$ 
or $\alpha$ for fixed nonlinearity parameter) is varied.  A mind-boggling 
multitude of bifurcations leading to the onset of chaos and 
further bifurcations occurs in the parameter space.  The details are 
further described in Sec. 4.
The above type of numerical analysis can be used in different ways to 
extract information about the dynamics.  Some of them are the following.
\begin{itemize}
\begin{enumerate}
\item
Trajectory plot ($(x$-$t)$-plot)
\item
Phase portrait: $(x$-$\dot x)$-projection of the trajectories from 
   three-dimen\-sional $(x,\dot x, t)$-space.
\item
Poincar\'{e} map: Stroboscopic (snapshot) portrait of the trajectory
   at every period $T (= \frac {2\pi}{\omega})$.
\item
Power spectrum: $ P(\hat \omega) = \left |\frac{1}{2\pi} 
   \int\limits_{-\infty}^{\infty} x(t) \exp{(-i\hat\omega t)}\right|^2$.  
\item
Bifurcation diagram: Plot of Poincar\'{e} points of the solution    
   $x(t)$ vs control parameter.
\item
Lyapunov spectrum: Plot of maximal Lyapunov exponent vs control
   parameter.
\end{enumerate}
\end{itemize}
Besides these, one often uses correlation functions, basins of
attractors and so on.

What are the kinds of behaviours a dissipative nonlinear system
of driven type typically exhibits and how do they occur?  The common 
types of behaviours are the following: 
\begin{itemize}
\begin{enumerate}
\item 
Fixed points
\item 
Periodic response of period $\omega$ of the input
\item 
Subharmonic response $(\Omega = \frac{\omega}{n}$, $n$ integer)
\item 
Superharmonic response $(\Omega = n\omega)$
\item 
Almost periodic (quasiperiodic) response
   (consisting of periodic components whose frequencies are
    incommensurable)
\item 
Chaotic response (nonperiodic and sensitively dependent on
   initial conditions)
\end{enumerate}
\end{itemize}

Some typical attractors are illustrated in Figs. 3.
There are typical routes through which periodic and chaotic responses
can arise as the control parameter varies.  Three most predominant
routes are the following:
\begin{itemize}
\begin{enumerate}
\item 
Feigenbaum's period doubling scenario.
\item 
Ruelle-Takens-Newhouse scenario of a quasiperiodic route.
\item 
Pomeau-Manneville scenario of an intermittency route.
\end{enumerate}
\end{itemize}
    Besides these three prominent routes, there are many other less 
common routes, such as period adding sequence, equal-periodic bifurcations
and so on, which occur occasionally in dissipative systems.

\subsection {Linear and nonlinear electronic circuits [6,8]}
Besides numerical and analytic studies of nonlinear oscillators there 
is another useful way of investigating them, namely by analog simulation
through suitable electronic circuits.  Such studies help in a dramatic
way to scan the parameter space quickly and to avoid long transients
as in the case of numerical studies.  The construction of such circuits
is fairly easy in typical cases, giving rise to quick responses as the control 
parameter is varied.

From another point of view, a variety of nonlinear electronic
circuits consisting of nonlinear physical devices such as nonlinear diodes,
capacitors, inductors and so on or devices constructed with
piecewise-linear
circuit elements have been utilized in recent times as model
nonlinear systems in their own right to
explore different aspects of chaotic dynamics.  They further
give rise to interesting technological applications.

\subsubsection {Linear circuits}
Linear circuits consist of only linear elements such as linear resistors,
capacitors or inductors besides the source.  In fact, the ubiquitous $LCR$
circuit (Fig. 4) can be considered as a prototype analog simulation of
the damped and
driven linear oscillator (8).  The $LCR$ circuit of Fig. 4 consists of a linear
resistor, a linear inductor, a linear capacitor and a time-dependent voltage
source, $ f(t) = F_s \sin\omega_s t$.  Applying Kirchhoff's voltage law to
the circuit of Fig. 4, we obtain
$$ L \frac {di_L}{dt} + Ri_L + v = F_s \sin\omega_s t, \eqno(11)$$
with $i_L(0) = i_0$, $v(0) = v_0$, where $i_L$ is the current along the 
inductor, $v$ is the voltage across the capacitor, $L$ is the linear inductance
and $R$ is the linear resistor.

Now substituting $i_L = C \frac {dv}{dt}$ into Eq. (11) and rearranging one
obtains the linear second-order inhomogeneous ODE with constant coefficients,
$$ \frac{d^2v}{dt^2} + \frac {R}{L} \frac{dv}{dt} + \frac{1}{LC} v =
\frac{F_s}{LC} \sin\omega_s t.  \eqno(12)$$
Obviously Eq. (12) is analogous to (6) with appropriate identifications,
and its solution can be immediately written down in the form (7) corresponding
to limit cycle oscillation with frequency $\omega = \omega_s$.  The actual
experimental result for the wave form $v(t)$ and the phase portrait in the 
$(v$-$i_L)$-plane is shown in Fig. 5, which is seen to mimic the 
solution of (6) given in Fig. 1(c).

\subsubsection {Nonlinear circuits [9]}
\leftline{\bf a) analog simulation}

As mentioned above, nonlinear oscillator equations can be analog simulated
by the use of suitable operational amplifier modules and multipliers.  Normally
op-amps are used as integrators, differentiators, sign-changers, adders and so on.
The mathematical operations are generally obtained either by using an op-amp
individually with resistor feedback, or an op-amp with capacitor
feedback.  Similarly, product
and division operations can be performed by multiplier and divider chips
such as AD532 or AD534.  One such analog simulation circuit constructed with 
adding integrators, scale changers and multipliers allows
us to study the chaotic
dynamics of the cubic Duffing oscillator equation (8).  Its schematic diagram
is given in Fig. 6.
\vskip 8pt
\leftline{\bf b) Piecewise-linear circuits: MLC circuit [10]}

Considering the familiar $LCR$ circuit given in Fig. 4, let us add a nonlinear
resistor, the so called Chua's diode characterized by a three segment negative
resistance characteristic curve as given in Fig. 7.  The modified circuit
is given in Fig. 8.  This simple non-autonomous circuit was first reported
by Murali, Lakshmanan and Chua [10].  The equations of motion governing
this circuit are 
$$ C \frac{dv}{dt} = i_L - g(v)  \eqno(13a)$$
$$ L\frac{di_L}{dt} = -v - Ri_L + F_s \sin\omega_st, \eqno(13b)$$
where $g(v) =G_bv+{1 \over 2}(G_a-G_b)[|v+B_p|-|v-B_p|]$, and $G_a$,
$G_b$ and $B_p$ are circuit parameters.
It can also be considered as a prototype of the chaotic nonlinear oscillators.

\section{Some ubiquitous damped and driven nonlinear oscillators [6]}
\subsection{Duffing oscillator}

One of the widely studied damped and driven nonlinear oscillators is the
Duffing oscillator given by Eq. (8). It can be considered to represent a
particle (of unit mass) moving
in a potential well
$$ V = \frac {1}{2} \omega_0^2 x^2 + \frac {\beta}{4} x^4, \eqno(15)$$
subject to viscous damping and external periodic force.  
Here one may distinguish three cases.

\begin{itemize}
\begin{enumerate}
\item 
Double-well (twin-well) potential $\omega_0^2 <0, \beta >0$
\item 
Single-well potential $\omega_0^2 >0, \beta >0$
\item 
Double-hump potential: $\omega_0^2 >0, \beta <0$.
\end{enumerate}
\end{itemize}

Each one of the above three cases is an important model for many 
physical problems.  The double-well potential represents the oscillations
of a buckled beam under a periodic force, plasma oscillations and so on.
Similarly, the single-well oscillator is a standard model of a particle
oscillating under viscous damping and a periodic force, while a double-hump
potential well is a model for weakly pinned charge density waves in
anisotropic solids and superionic conductors.

Now Eq. (8) has an obvious sealing property which allows one to fix two
of the five parameters $ \alpha$, $\omega_0^2$, $\beta $, $f$ and $\omega$.
From an experimental point of view it is more convenient to fix the
spring forces $\omega_0^2$ and $\beta$ and vary the parameters in the 
$(\alpha $-$ \omega $-$f)$-space.  From a practical point of view one normally
studies behaviour in the $f$-space or $(f$-$\omega)$-space by fixing the
remaining parameters as well.  In the next section (Sec. 4), we will 
briefly discuss the chaotic dynamics of all the three potential wells 
of Eq. (8).
\subsection{BVP and DVP oscillators}

The nerve impulse propagation along the axon of a neuron can be studied
by combining the equations for an excitable membrane with the differential
equations for an electrical core conductor cable, assuming the axon to be
an infinitely long cylinder.  A well known approximation of FitzHugh and
Nagumo to describe the propagation of voltage pulses $V(x,t)$ along the
membranes of nerve cells is the set of coupled PDEs
$$ V_{xx} - V_t = F(V) + R - I,  \eqno(16a) $$
$$ R_t = c(V+a-bR), \eqno(16b) $$
where $R(x,t)$ is the recovery variable, $I$ the external stimulus and
$a$, $b$, $c$ are related to the membrane radius, specific resistivity of the fluid inside
the membrane and temperature factor respectively.  Here suffixes stand
for partial differentiation.

When the spatial variation of $V$, namely $V_{xx}$, is negligible, Eq. (16)
reduces to a set of two coupled first-order ODEs known as the Bonhoeffer-van
der Pol oscillator system,
$$ \dot V = V - \frac{V^3}{3} - R + I \eqno (17a)$$
$$ \dot R = c(V+a-bR),  \,\,\, \left (. = \frac{d}{dt}
\right )\eqno(17b)$$
with $ F(V) = -V + \frac{V^3}{3}$.  Normally the constants in Eq. (16) satisfy
the inequalities $b<1$ and $3a+2b>3$, though from a purely mathematical
point of view this need not be insisted upon.  Then with a periodic (ac) 
applied membrane current $A_1 \cos\omega t$ and a (dc) bias $A_0$, the BVP
oscillator equation becomes
$$ \dot V = V - \frac{V^3}{3} - R + A_0 + A_1 \cos\omega t, \eqno(18a)$$
$$ \dot R = c (V+a-bR). \eqno (18b) $$
Again system (18) shows a very rich variey of bifurcations and chaos
phenomena as discussed in Sec. 5 below.

Further, Eqs. (18) can be rewritten as a single second-order differential
equation by differentiating (18a) with respect to time and using (18b) for $R$,
$$ \ddot V-(1-bc)\left \{1-\frac{V^2}{1-bc}\right \}\dot V -c(b-1)V + \frac{bc}{3}V^3
=c(A_0 b-a)+A_1 \cos(\omega t + \phi),  \eqno(19) $$
where $\phi = \tan^{-1} \frac{\omega}{bc}$.  Using the transformation
$ x = (1-bc)^{-(1/2)}V$, $t\longrightarrow t' = t+\frac{\phi}{\omega}$,
Eq. (19) can be rewritten as
$$ \ddot x+p(x^2-1)\dot x+\omega_0^2x+\beta x^3=f_0+f_1\cos\omega t, \eqno(20)$$
where 
$$p = (1-bc), \,\,\, \omega_0^2 = c(1-b), $$
$$ \beta = bc\frac{(1-bc)}{3},\,\,\, f_0 = c\frac{(A_0b-a)}{\sqrt {1-bc}},$$
$$ f_1 = \frac{A_1}{\sqrt {1-bc}}. \eqno(21)  $$ 
Eq. (20) or its rescaled form
$$ \ddot x+p(kx^2+g)\dot x + \omega_0^2 x +\beta x^3 = f_0 +f_1 \cos\omega t \eqno(22) $$
is the Duffing-van der Pol equation.  In the limit k=0, we have the Duffing
equation discussed above (with $f_0=0$), and in the case $\beta = 0$ ($g=-1$, $k=1$) 
we have the driven van der Pol equation.  Eq. (22) again exhibits a very rich
variety of bifurcations and chaos phenomena, including quasiperiodicity,
phase lockings and so on, depending on whether the potential $ V=\frac{1}{2}
\omega_0^2 x^2 + \frac {\beta x^4}{4}$ is i) a double well, ii) a single
well
or iii) a double hump.  We will discuss some of the details in Sec. 5. 
\section{Duffing oscillator: Bifurcations and chaos}

As noted earlier, the Duffing oscillator (Eq. (8)) dynamics can be described
by a phase diagram in a three-parameter ($\alpha$-$f$-$\omega$)-space.  In the
following, we will concentrate on the $f$- or $\omega$- or ($f$-$\omega$)-parameter 
space because of practical considerations.
\subsection {The double-well oscillator $ (\omega_0^2<0, \beta>0)$}

Considering a particle in a double-well potential $ V = -1/2 |\omega_0^2| x^2 +
\beta x^4 $ with a base vibrating periodically, we notice that there are three
equlibrium points in the force-free case $(f=0)$,  corresponding to $\omega_0^2 x + 
\beta x^3 = 0$.  They are the stable fixed points (centres)
$$ x_{1,2}^s = \pm \sqrt{ \frac{|\omega_0^2|}{\beta}} \eqno(23) $$
and an unstable fixed point which is a saddle, 
$$ x_0^u = 0. \eqno(24) $$
Then the oscillations about the two stable equilibrium points are
given (after a redefinition of $x$) by
$$ \ddot x + \alpha\dot x + 2|\omega_0^2|x \mp 3\sqrt{|\omega_0^2| \beta}
\,\,x^2 +\beta x^3 =f \sin\omega t.  \eqno(25) $$

\subsubsection {Period doubling scenario in $f$-space}

Choosing the initial condition such that the system approaches
the left equilibrium point $(x_1 = -1.0)$ asymptotically as example,
and choosing the parameters
as $\omega_0^2 = -1.0 $, $\beta = 1.0$, $\alpha = 0.5$, $\omega = 1.0$, 
we can solve Eq. (25) numerically and obtain the following picture.
To start with we realize from the theory of nonlinear oscillations that the
system can exhibit two types of periodic motions:
\begin{itemize}
\begin{enumerate}
\item
Small orbits $(SOs)$: Oscillations around the stable fixed point $x_1$.
\item
Large orbits $(LOs)$: Large amplitude oscillations that encircle all the
three equlibrium points $x_1,x_2$ and $x_0$.
\end{enumerate}
\end{itemize}
Then the following picture arises.

{\bf A. Small orbit oscillations }

As $f$ is increased slightly from zero, a stable period $T$ limit cycle about
the left equilibrium point $x_1 = -1.0 $ occurs, which persists up to $f=0.34$.  Then
a period doubling scenerio starts at $f = 0.35$ with a $2T$ periodic cycle, followed
by a $4T$ periodic cycle at $f=0.358$, etc. This cascade of bifurcations accumulates
at $f = f_c = 0.361$, the sequence converging geometrically at the rate
known as the Feigenbaum ratio.  Beyond $f=f_c$, one finds
chaotic orbits occurring where the motion is highly dependent on the initial 
conditions. This can be checked by the various characterizations 
discussed previously.  These motions are illustrated in Figs. (9a-c).

{\bf B. Hopping Oscillations} 

When $f=0.37$, the onset of a crisis occurs, in which
the chaotic motion that had been confined to the left well suddenly expands to 
cover the right well also, in a hopping cross-well chaotic motion.

{\bf C. Large orbit oscillations}

For still larger values of $f$, the chaotic state suddenly disappears
and is replaced by a
stable period-$mT$ oscillation, overcoming the central barrier, extending
over both the valleys and encompassing
all the equilibrium points. (Fig. 9f, g).

Fuller details are given in Table 1 and the bifurcation diagram in Fig. 10.

{\bf D. Antimonotonicity}

In addition to the periodic window resulting from the crisis, there is also
another possibility due to antimonotonicity.  Here one finds that periodic
orbits are not only created through period doubling cascades but are also 
destroyed through reverse period doubling when a control parameter is
monotonically increased in any neighborhood of a homoclinic tangency 
value.  Two types of such antimonotonicity are shown in Fig. 11.  Fig. 11(a)
represents a period-3 bubble in the range $0.53 \leq f \leq 0.62$, while 
Fig. 11(b) shows a reverse-period bubble in the region $0.43\leq f \leq 0.48$.
\subsubsection {Phase-diagram in $(f$-$\omega)$-space}

To appreciate the rich variety of bifurcations, let us look at the
$(f$-$\omega)$-phase
diagram (Fig. 12) given by Szemplinska-Stupnicka and Rudowski [11] in the
frequency range $ 0.25 <\omega <1.1$, fixing $ \alpha = 0.1$,
$|\omega_0|^2 = 0.5$ and $\beta = 0.5$.  One observes
that $SO$, symmetric $LO$, unsymmetric $LO$ and
cross-well chaotic or stable attractors coexist.  The $SO$ motions occur
within the whole $(f$-$\omega)$-plane except for two $V$-shaped regions,
one with a cusp
at $\omega = 0.8$ and $f = F_2 $ at the principal resonance region, the
other cusp at $\omega  = 0.4$ , and $f = 0.14$, that is the superharmonic zone.
Inside the two $V$-shaped regions the system can exhibit cross-well chaotic
(or regular) motion.  Also, in some regions, two different attractors coexist
while, in the others, a single steady state (globally stable) motion can be observed.
Typical orbits: the form of the various single and coexisting states
denoted in Fig. 12 by numbers $1,2, \ldots, 20$ are shown in Fig. 13.  Regular
attractors are illustrated by their phase portraits and chaotic attractors
by Poincar\'e maps.  They include all the types of motions discussed above.
\subsection {Single well and double hump potentials}

Considering the single-well potential, one observes hysteresis
in the primary resonance region, harmonic resonances and 
symmetry breaking orbits for a range of low values of $f$.
For sufficiently high values of $f$,
chaos occurs through a period-doubling route.
Similarly, in the double-hump case, one finds that one symmetric attractor
bifurcates into two mutually symmetric attractors.  Both attractors then
undergo period-doubling cascades to chaos, where windows of both even and 
odd periods occur.  Then both chaotic attractors broaden and merge to
form a single attractor, ultimately leading to a boundary crisis.  For further
details see Ref. [6].
\subsection {analog simulation and experimental verification}

The Duffing oscillator (8) can be easily analog-simulated using op-amps
and four-quadrant multipliers.  Replacing $\dot x(t) = \int \ddot x dt$, 
$x(t) = \int \dot x dt$, the equivalent schematic simulation circuit is given in
Fig. 6.  Here blocks $IC_1$ and $IC_2$ represent two integrators, $IC_3$ is an 
inverter (sign-changer), and $M_1$ and $M_2$ are four-quadrant multipliers.
The experimental results for the double-well case are given in Figs. 14, and
they agree with the theoretical numerical analysis.  For further
details see [6] again.
\section {Dynamics of the BVP and DVP oscillators [6] and MLC Circuit}

Let us now quickly review the chaotic dynamics 
and bifurcations of the BVP and DVP oscillators discussed
in Sec. 3.2.  Apart from the period-doubling route to chaos, one also observes
the presence of phase-lockings and devil's staircase structures, quasiperiodic
and intermittency routes to chaos predominantly.  The nonlinear dynamics of the
MLC circuit are also discussed.  We shall therefore give only a very brief
account of these structures.
\subsection {BVP oscillator}

Considering Eq. (18), the equilibrium points in the absence of external
stimuli $(A_0 = A_1=0)$ are the roots of the equations
$$ V^3 - qV - p = 0, \eqno(26a) $$
$$ V + a - bR = 0, \eqno(26b) $$
where $ p = \frac{-3a}{b}$ and $q = 3 \frac{(b-1)}{b}$.  Eq. (26a)
has three real roots for $27p^2 >q^3$ and one real root for $27p^2 <q^3$. 
The linear stability analysis of Eqs. (26) shows that for a fixed point ($V_0$, $R_0$)
to be stable, the necessary and sufficient condition is 
$$ (V_0^2+bc-1) >0 $$ and $$ c(1+bV_0^2-b)>0. \eqno(27a) $$
Thus for suitable choices of parameters one can have one stable focus or
one saddle and two stable foci.

Proceeding along the same lines, with the inclusion of constant bias
($A_1=0$, $A_0>0$), one can find fixed points by simply modifying
$p$ to $ p = 3(A_0 -a/b) $ in Eqs. (26).  It is of particular interest to check 
whether the fixed points undergo a Hopf bifurcation.  This can be easily 
verified by checking the nature of changes in the stability property of the
fixed points and the parametric choices for which the eigenvalues cross the
imaginary axis.  For the present problem, one can easily check that a Hopf
bifurcation occurs at
$$ A_{0\pm}=\frac{V_{0\pm}^3}{3}-\frac{b-1}{b}V_{0\pm}+\frac{a}{b} \eqno(27b) $$
One checks that for $ A_0<A_{0-}$ and $A_0>A_{0+}$ the fixed points are 
asymptotically stable, being either a node or a spiral point.  
For $ A_0 \in (A_{0-}, A_{0+})$, the eigenvalues have positive real
parts so that the fixed point is unstable and one may expect a stable 
limit cycle in this range.

Similarly when only a periodic input current $(A_1 \neq 0, A_0 = 0)$ is 
present, the system, as in the case of the Duffing oscillator, undergoes
period-doubling bifurcations to chaos, followed by windows, intermittency 
and so on.  The bifurcation diagram given in Fig. 15 describes briefly
the various bifurcations admitted by this system for the parametric 
choice $a = 0.7$, $b = 0.8$, and $c = 0.1.$

Now if we consider the situation in which both external d.c. and 
periodic stimuli are present, the system enters into an interesting 
phenomenon called phase-locking or mode-locking.  Here the natural frequency
of the oscillator $(\nu_n)$ and the frequency of the driving force 
get locked.  When one calculates the winding number $ W = \frac{\nu_d}{\nu_n}$
as a function of the d.c. current $A_0 $ (which will give rise to a different
$\nu_n$), the corresponding schematic diagram has a devil's staircase
structure.  A typical example for the choice $A_1 = 0.74$ is shown in Fig. 16.

\subsection {DVP oscillator}

One can consider three different cases for the DVP oscillator (22), as in the
case of the Duffing oscillator, namely i) double well, (ii) single well 
and (iii) double hump.  They exhibit many complicated bifurcation phenomena
including period-doubling, hysteresis,  quasiperiodicity, phase-locking and 
so on.  Some interesting features of the double-hump DVO oscillator include

i)    the existence of a Farey sequence,

ii)   double hysteresis loops,

iii)  a swallow-tailed bifurcation structure of subharmonic locking,

\noindent and so on.  For more details see Ref. [6].

\subsection {Chaotic dynamics of the MLC circuit [10]}
The simple second-order nonautonomous dissipative nonlinear circuit 
consisting of Chua's diode as its only nonlinear element, suggested
by Murali, Lakshmanan and Chua $(MLC)$, can exhibit a rich variety of
bifurcations and chaos phenomena.  The circuit diagram and its state
equations have already been introduced in Sec. 2.3.2.  Rescaling Eq. (13) with
$ v = x B_p$, $i_L = G.y.B_p$, $G=1/R$, $\omega = \frac {\Omega C}{G}$ and
$t = TCG$, and then redefining $T$ as $t$, we have the following set of 
normalized equations,
$$\dot x=y-h(x),\,\,\, \dot y = -\beta y -\nu \beta y - \beta x + 
f\sin\omega t,  \eqno(28a) $$
where $\beta=\frac{C}{LG^2}$, $\nu=GR$, and $f = \frac{Fs}{\beta_p}$.
Here $h(x)$ stands for
$$h(x) =  \left\{\matrix {
        \hfill bx+a-b, & x\geq 1   \cr
        \hfill     ax, & |x|\leq 1 \cr
        \hfill bx-a+b, & x\leq -1 . \cr
}\right. \eqno(28b) $$
Here $a = \frac{Ga}{G}$, $b = {G_b}{G}$.
Now the dynamics of Eq. (28) depend on the parameters $\nu$, $\beta$, $a$, $b$, 
$\omega$ and $f$.  The experimental circuit parameters used in the actual 
circuit correspond to $\beta=1$, $\nu=0.015$, $a=-1.02$, $b= -0.55$ and $\omega=0.75$.
It is straightforward to establish that there exists a unique equilibrium $(x_0,y_0)$
for Eq. (28) in each of the following three regions,
$$
\matrix{
\hfill D_1 = \{(x,y)|\,\, x\geq 1\}:       & P^+ = (-k_1, -k_2) \cr
\hfill D_0 = \{(x,y)|\,\, |x|\leq 1\}:     & O = (0,0)      \cr
\hfill D_{-1} = \{(x,y)| \,\, x\leq -1 \}: & P^- = (k_1,k_2)
}
$$
where $k_1 = [(\sigma(a-b))/(p+\sigma b)]$, $k_2 = 
[\beta (b-a)/(\beta+\sigma b)]$ and $\sigma = \beta (1+\nu)$.
In each of the regions $D_1, D_0, D_{-1}$, Eq. (28) is linear.  One can
easily check by a stability analysis that $O \in D_0 $ is a 
hyperbolic (saddle) fixed point, while $P^+$ and $P^-$ are stable spiral 
fixed points.  Depending on the initial conditions, any of them can be
observed.  As the forcing signal $f$ is increased ($f>0$), the stable fixed
points undergo a Hopf bifurcation.  A further increase in $f$ leads to period
doubling bifurcations and finally to chaos.  The corresponding numerical
and experimental results are given in Fig. 17.
\newpage                                                                                                                                                       
\section {Dictionary of Bifurcation}
The various oscillator systems discussed in the previous sections
exhibited a variety of oscillatory behaviours including fixed
points, resonant periodic oscillations, period-doubled oscillations,
quasiperiodic and chaotic motions.  In particular, the various bifurcation
diagrams clearly show that all these systems exhibit qualitative (and
quantitative) changes at critical values of the control parameters 
as they are varied smoothly.  These correspond to the  different
bifurcations exhibited by the systems.
The question is how to understand the existence of such bifurcations
and the mechanism by which they arise.  One possible approach is to
look for the local stability properties of solutions in the neighbourhood
of the critical parameter values at which bifurcations occur.  However, since
no exact solutions are available for the above systems, one can prepare
a dictionary of bifurcations [12] that occur in simple and interesting
low dimensional nonlinear systems, and then use them as reference systems
for the identification of the bifurcations in the various nonlinear oscillators.
In Tables 2 and 3 we give their salient features which are
self-explanatory.

\section {Analytic approaches}
The various types of bifurcations exhibited by a typical nonlinear
oscillator can be classified by using the dictionary of bifurcation
phenomena like the ones discussed in the earlier section.  However,
some approximate analysis can be performed analytically by constructing
perturbative solutions and studying their stability properties.  Moreover
the analytic structure of solutions in the complex $t$-plane can throw
much light on the integrability and nonintegrability and chaotic
behaviour of dynamical systems.  We will discuss some features of these
aspects in this section.

\subsection{Theoretical analysis of large orbit $T$-periodic solution
in the double-well Duffing oscillator}
The numerical analysis presented in Sec. 4 for the double-well Duffing
oscillator conclusively showed that it admits a large orbit ($LO$) $T$-periodic
solution for a wide range of drive frequencies.  One might then expect a
first-order approximate solution whose stability analysis might give a
good estimation of the parameter domain in which it occurs.  So
we shall look for a T-periodic solution of the oscillator equation [11],
$$ \frac{d^2x}{dt^2}+\alpha \frac{dx}{dt}-|\omega_0^2|x+\beta x^3 =
f\cos\omega t,   \eqno(29)$$
which is close to a harmonic function of time as
$$x(0) = A(0) \cos(\omega t+ \phi), \eqno(30) $$
where $A(0)$ and $\phi$ are to be determined.  For this purpose we
transform (29) into the form
$$ \ddot x +\omega ^2 x + \varepsilon f(x,\dot x, t) = 0 \eqno(31) $$
as
$$ \ddot x +\omega^2 x + \mu (\bar\alpha \dot x + x \sigma -
\Omega^2 x - |\bar\omega_0|^2 x + \bar\beta x^3 - \bar f\cos\omega 
t)=0, \eqno(32) $$
where
$$ \mu \bar\alpha = \alpha, \,\,\,  \mu \sigma = \mu \Omega^2 -\omega^2, \,\,\,
\mu \bar f =f, \,\,\, \mu |\bar\omega_0|^2 = |\omega_0|^2,  
\mu\bar\beta = \beta. \eqno(33) $$
Here the quantity $\Omega$ has been introduced from the form of the
harmonic solution $x = a \cos\Omega t$, $\Omega^2 = -|\omega_0^2| +
3/4 \beta a^2>0$.
Now one can look for a periodic solution of (32) in the form 
$$x(t,\mu) = x_0(T_0,T_1, \ldots)+\mu x_1(T_0,T_1, \ldots), \eqno(34) $$
where the multiple scales $T_0 = t$, $ T_1=\mu t$, $\ldots $
and $\frac{d}{dt} = \sum\limits_{n=0}^{\infty}\mu^n D_n$,  $D_n=
\frac{\partial} {\partial T_n}$ so that we have the system of equations
$$ D_0^2 x_0 + \omega^2 x_0 = 0, \eqno (35a) $$
$$D_0^2 x_1 + \omega^2 x_1 = -2D_0 D_1x_0 - \bar\alpha D_0x_0
-(\sigma - \Omega^2-|\bar\omega_0)|^2) x_0 - \bar\beta x_0^3 + 
\bar f \cos\omega t,  \eqno(35b) $$
\hskip 5 pt etc. 
The general solution of (35a) is
$$ x_0 = A_1(T_1) \exp i\omega T_0 + c.c., \eqno(36) $$
where $A_1$ is, in general, complex.  Using (36) in (35b), in order to
eliminate the secular terms in the solution of $x_1$, one requires that
$$-[2i\omega (D_1A_1 + \bar\alpha/2 A_1) + (\sigma - \Omega^2 -
|\bar \omega_0|^2)A_1 + 3\beta A_1^2 A_1^*] + \bar f/2 = 0. \eqno(37)$$
Making the substitution
$$ A_1(T_1) = (1/2) a(T_1) \exp i \phi (T_1) \eqno(38) $$
and equating real and imaginary parts for steady state conditions
with $\Omega^2 \approx  -|\omega_0^2| + 3/4 \beta a^2$, 
we obtain
$$ a =\frac{f}{\sqrt{(\Omega^2 -\omega^2)^2 + \alpha^2\omega^2}},\,\,\,\,
\tan\phi = \frac{-\alpha\omega}{(\Omega^2(a)-\omega^2)}. \eqno(39)$$

Finally, the first-order correction becomes
$$\mu x_1(t) = A_3 \cos 3(\omega t + \phi), \,\,\,\,
A_3 = \frac{\mu\bar\beta a^3} {32\omega^2} = \frac{\beta a^3}{32\omega^2},
\eqno(40) $$
so that the $LO$ solution becomes

 $$ x(t) = a\cos(\omega t + \phi) + \frac{a^3} {32\omega^2} 
 \cos 3(\omega t+\phi) \equiv x(t+T). \eqno(41) $$

 \subsubsection {Stability of $LO$ solution,}
\noindent  {\bf A. Soft mode instability}

Considering small disturbances to the amplitude and phase of the solution   
(41) leading to
$$ \tilde x(t) = (a+\delta a) \cos (\omega t +\phi + \delta \phi) + 
\frac{(a+\delta a)^3}{64 \omega^2} \cos 3(\omega t +\phi +
\delta \phi), \eqno(42) $$
one obtains the linear eigenvalue problem 
$$\pmatrix{ 
\delta \dot a \cr
\delta \dot \phi } =
\pmatrix{
-{\alpha \over 2} & -{\alpha \over 2\omega}(\Omega^2-\omega^2) \cr
{1 \over 2\omega a}(\Omega^2-\omega^2) & -{\alpha \over 2}}
\pmatrix{\delta a \cr \delta \phi}. \eqno (43) $$
Then the linear stability of the solution (41) depends upon the 
($2 \times 2$) 
matrix in the right hand side of (43) from which a necessary
criterion for the stability of the solution can be obtained (see 
Fig. 20).

\noindent {\bf B. Symmetry-breaking instability}

Considering more general instabilities, $ \tilde x(t) = x(t) + \delta x(t)$,
one can deduce a Hill type eigenvalue problem,
$$\delta \ddot x + \alpha \delta \dot x + (\lambda_0 + \sum_{n=2,4,6}
\lambda_n \cos n \omega t)\delta x = 0, \eqno(44)$$
where $\lambda_0$, $\lambda_2$, $\lambda_4$, $\lambda_6$ are appropriate
coefficients.  One can check that while there is no period-doubling
instability, symmetry-breaking instability can occur here.
This may be verified by assuming a solution of the form
$$\delta x(t) = e^{\epsilon_1 t} (b_0 + b_{21} \cos 2\omega t + b_{22} 
\sin 2\omega t), \eqno(45)$$
and finding the values of $\epsilon_1$.  The corresponding curve
$\omega_{SB}(a)$ is given in Fig. 20.
One finds that the curves $ \omega_B (a)$  and $\omega_{SB}(a)$ determine
the region of existence of $LO$ to a reasonable approximation.

\subsection {Stability analysis of $SO$ motion}
The numerical analysis of the double-well Duffing oscillator discussed earlier
clearly shows the presence of $SO$ periodic motion confined to one of the
wells.  One can therefore again look for $T$- periodic solution in the 
right well for example and apply multiple scale perturbation theory.
Comparing with Eq. (29), we consider the equation
$$\ddot x+ \omega^2 x + \mu \bar \omega_1 x^2 + \mu^2(\bar \alpha \dot x+
\bar \omega_3 x^3 + \sigma x - \bar f \cos\omega t) = 0,$$
where
$$\mu \bar \omega_1 = 3/2, \\ \mu^2 \bar \omega_3 = 1/2,\\ \mu^2 \bar \alpha
= \alpha,\\ \sigma \mu^2 = 1- \omega^2, $$
$$ \mu^2 \bar f = f, \mu \ll 1.  \eqno(46)$$
Proceeding as in the previous case one obtains the second-order approximate
solution
$$x(t) = a \cos (\omega t + \phi) - 3/4 a^2 + 1/4 a^2 \cos 2(\omega t + \phi)
+ a^3/16 \cos 3(\omega t + \phi). \eqno(47)$$
Considering, as in the case of $LO$ periodic solutions, the stability of the
$SO$ periodic solution (47), one can analyse the first-order instability 
as well as period-doubling instability.  As a result one can conclude
that the approximate $T$-periodic solution is unstable within a region of the
driving frequency $ \omega_A < \omega < \omega_{PD}$ and that no other $SO$
stable solution is possible.  The above result is found to give rise to a
surprisingly good estimation of the system's critical parameter values 
for which cross-well chaos really occurs.  For more details, see Ref. [11].

\subsection{Analytic structure of the Duffing oscillator}
An altogether different approach to investigating the nonintegrability
aspects of the Duffing oscillator and other dynamical systems is the 
study of the analytic structure of singularities of the solutions in the
complex time plane, known in the literature as the Painlev\'{e} singularity
structure analysis [see for example Refs. 13-15].  Originally this method
was applied successfully to isolate the integrable cases of the rigid body
motion in a gravitational field.  Its recent renaissance started with
the discovery of soliton systems (see the lectures of Prof. Ablowitz
in this School).
One essentially tries to isolate the parameters of the nonlinear differential
equations underlying the dynamical system for which the solution is free
from movable critical points (movable branch points and essential 
singularities) so that the solution is meromorphic in the complex time
plane.  For nonintegrable systems, the singularities are often located
in a complicated pattern, giving an indication that a sufficient number of
analytic integrals is lacking.
Consider for example the unforced, undamped (free) Duffing oscillator,
$$ \ddot x +\omega_0^2x + \beta x^3 = 0. \eqno(48)$$
The solution to (48) can be given in terms of Jacobian elliptic functions,
$$ x(t) = A cn (\Omega t + \delta), \\ \Omega = {\sqrt{ \omega^2 +
\beta A^2}}, \eqno(49)$$
where the modulus $k ={\sqrt{ \beta A^2/2 (\omega_0^2 + \beta A^2)}}$.  One
knows that the elliptic functions are meromorphic and doubly periodic. The 
only singularities they possess in the complex $t$-plane are poles of order $1$ 
at the lattice points,
$$ (\Omega t + \delta) = 2nK + (2m+1)iK', \\\ n,m  \in Z. \eqno(50)$$
We also note that despite the existence of these poles, the
system possesses the analytic integral, $ E= \frac {1}{2} \omega_0^2 A^2+
\frac {1}{4} \beta A^4 $, in the complex $t$-plane and the system is integrable.
A typical pattern of the lattice of singularities of the solution of
(48) is given in Fig. 21a.
The situation becomes more complicated when damping as well as forcing
are present.  Typical singularity structures are shown for the case when
damping alone is present in Fig. 21b, and when forcing is also present
in Fig. 22.
One observes a progressive clustering of singularities off the real axis
as the forcing strength increases.  Actually, one can identify much more 
intricate structures through local analysis or numerical analysis.  
Around each singular point there is an infinite-sheeted, four-armed
structure of singularities.  For more details, see Refs. [13-15].
Similarly, infinite-sheeted structures are present in the case of the DVP
oscillator [16].

\section{Controlling Chaos}

Eventhough  the  presence  of  chaotic 
behaviour is generic and robust for suitable nonlinearities,
ranges of parameters and external forces, there are practical
situations where one wishes to avoid or control chaos so as to
improve the performance of the dynamical system.  Also, eventhough
chaos is sometimes useful as in a mixing  process or in heat
transfer, it is often unwanted or undesirable.  For example,
increased drag in flow systems, erratic fibrillations of heart
beats, extreme weather patterns and complicated circuit
oscillations are situations where chaos is harmful.  Clearly,
the ability to control chaos, that is to convert chaotic
oscillations into desired regular ones with a periodic time
dependence would be beneficial in working with a particular
system.  The possibility of purposeful selection and stabilization
of particular orbits in a normally chaotic system, using minimal,
predetermined efforts, provides a unique opportunity to maximize
the output of a dynamical system.  It is thus of great practical
importance to develop suitable control methods and to analyze
their efficacy.  Very recently much interest has been focussed on
this type of problems [6,17-19].

\subsection{Algorithms for controlling chaos: Feedback and non-feedback
methods}
Let us consider a general $n$-dimensional nonlinear dynamical system, 
$$ \dot X = \frac {dX} {dt} = F(X \ ; \ p \ ; \ t),  \eqno(51)$$
where $X = (x_1, x_2, x_3, ..., x_n)$ represents the $n$ state variables 
and $p$ is a  control  or  external  parameter. Let $X(t)$ be a chaotic    
solution of Eq. (51). Different  control
algorithms are essentially based on the fact that one would like
to effect the most minimal changes to the original system so
that it will not be grossly deformed.  From this point of view,
controlling methods or algorithms can be broadly classified into
two categories:

    (i) feedback methods, and

   (ii) non-feedback algorithms.

    Feedback methods essentially make use of  the  intrinsic 
properties of chaotic  systems,  including  their  sensitivity  to 
initial conditions, to stabilize orbits already  existing  in  the 
systems.  Some of the prominent methods are the following. 
\begin{itemize}
\begin{enumerate}
\item  Adaptive control algorithm
\item  Ott-Grebogi-Yorke (OGY) method of stabilizing unstable 
       periodic orbits
\item  Singer {\it et al.}'s method of stabilizing unstable periodic orbits
\item  Various control engineering approaches.
\end{enumerate}
\end{itemize}

In contrast to feedback control techniques, non-feedback 
methods make use of a small perturbing  external  force 
such as a small  driving  force,  a  small  noise  term,  a  small 
constant bias or a  weak  modulation  to  some  system  parameter.  
These methods  modify  the  underlying  chaotic  dynamical  system 
weakly so that stable solutions appear. Some of the important controlling
methods of this type are the following. 

\begin{itemize}
\begin{enumerate}
\item  Parametric perturbation method
\item  Addition of a weak periodic signal, constant bias or noise
\item  Entrainment-open loop control
\item  Oscillator absorber method.
\end{enumerate}
\end{itemize}

    For more complete details of the various controlling methods, see for
example Refs. [6,17,18].

    Here is a typical example of adaptive control algorithm. We can control
the chaotic orbit $X_s$ = $(x_s,y_s)$ of the BVP oscillator by introducing
the following dynamics on the parameter $A_1$:

$$ \dot x = x-\frac{x^3}{3} - y + A_0+A_1 \cos\omega t, \eqno(52a)$$
$$ \dot y = c(x+a-by), \eqno(52b)$$
$$ \dot A_1 = -\epsilon[(x-x_s) - (y-y_s)],\\\\\ \epsilon<<1  \eqno(52c)$$
The actual conversion of the chaotic motion into period-$T$ and $2T$ limit
cycles in a typical case is shown in Fig. 23.  All the other methods
can be discussed in similar fashion.

\section{Synchronized Chaotic Systems and Secure Communication}          
Finally, we will discuss very briefly
another fascinating concept, namely, synchronization of chaos,
which is attracting considerable interest among chaos researchers
in recent times.  The possibility of two or more chaotic 
systems oscillating in a coherent and synchronized way is not an
obvious one.  One of the main features often associated with the
definition of chaotic behaviour is the sensitive dependence on
initial conditions.  Then one may conclude that synchronization is
not feasible in chaotic systems because it is not possible in
real systems either to reproduce exactly identical initial
conditions or exact specification of system parameters of two
similar systems. We may be able to build "nearly" identical
systems, but there is always an inevitable technological mismatch
and noise, impeding the exact reproduction of all the parameters and
initial conditions. Thus, even an infinitesimal deviation in any
one of the parameters will eventually result in the divergence of
nearby starting orbits.

          In this connection, the recent suggestion of Pecora and
Carrol [20] that it is possible to synchronize even chaotic
systems by introducing appropriate coupling between them has
revolutionized our understanding.  They have shown that if a
reproduced part (response or subsystem) of a chaotic system
responds to a chaotic signal from the full (drive) system, under
some conditions the signals in the response part would converge to
the corresponding signals in the drive system.  Such a possibility
is known as the synchronization of chaotic systems.  This idea
has been further generalized by cascading [20] the reproduced parts
or subsystems.  Here the considered chaotic system may be divided
into two subsystems, each of which will synchronize with the full
system  when  driven by the  appropriate chaotic signal.  By
arranging one of these synchronized subsystems to drive the other
subsystem, it is possible to produce synchronized signals in the
subsystems for every signal in the full drive system.

          Another innovative method which has been developed
recently for chaos synchronization is the method of one-way
coupling between two identical chaotic
systems [21,22].  In this configuration the response
chaotic system variables follow identically the drive chaotic
system variables for appropriate one-way coupling strength. 
Moreover, the behaviour of the response system is dependent on the
behaviour of drive system, but the drive system is not influenced
by the response system's behaviour.

          Thus by reproducing all the signals at the receiver
under the influence of a single chaotic signal from the drive, the
synchronized chaotic systems provide a rich mechanism  for  signal
design and generation, with potential applications to
communications and signal processing.  Further, one can use the
pair of cascading driven  subsystems or the identical response
system with one-way coupling term as attractor recognition devices
for chaotic attractors.  Hence, a single input signal will allow
recognition and reconstruction of the driving system and/or
rejection of non-matching device systems.  Another interesting
area of application would be cryptography.  For  example, the
non-driven part of the chaotic system produces signals, which  are
typically broad-band, noise-like and difficult to predict, and  so
they can be ideally utilized in various contexts for masking
information bearing waveforms.  They can also further be used as
modulating waveforms in spread spectrum techniques.

\subsection {Pecora and Carroll method}

          In this method, Pecora and Carroll have considered an
$n$-dimensional system governed by the state equation of the form
          
$$ \frac{dX}{dt} = f[X(t)],~X = (x_1,x_2, \cdots,x_n)^T. \eqno(53)$$

They divide the system into two parts in an arbitrary way as  $X  = 
(x_D, x_R)^T.$  The $D$ part is referred to as the  driving subsystem
variables and the $R$ part as the response subsystem variables.
Then Eq. (53) can be rewritten as
$$\dot x_D  = g(x_D,  x_R ), \,\,\,\,\,\,  (m{\rm -}dimensional) \eqno (54a) $$
$$\dot x_R  = h(x_D , x_R ), \,\,\,\,\,\,   (k{\rm -}dimensional)\eqno (54b) $$
where $x_D = (x_1, x_2 ,\cdots,x_m)^T$,    $x_R = (x_{m+1},x_{m+2},\cdots 
x_n)^T$, $g = [f_1(X), \cdots,$ $f_m(X)]^T$, $h = [f_{m+1}(X),\cdots, f_n(X)]^T$,  
and $m+k = n$.

          Pecora and Carroll suggested building an identical copy
of the response sub-system with variables $x'_R$  and  drive it with
the $x_D$  variables coming from  the  original  system.  In such a
model, we have the following compound system of equations:
$$  \dot x_D = g(x_D , x_R ), \,\,\,\,\,\,   (m{\rm -}dimensional){\rm -}drive \eqno(55a)$$           
$$  \dot x_R = h(x_D , x_R ), \,\,\,\,\,\,   (k{\rm -}dimensional){\rm -}drive \eqno(55b)$$
$$  \dot x'_R = h(x_D,x'_R), \,\,\,\,\, \,\,\,\,(k{\rm -}dimensional){\rm -}response.
    \eqno(55c)$$
Under the right conditions, as time elapses the $x'_R$ variables will
converge asymptotically to the $x_R$  variables and continue to remain
in step with the instantaneous values of $x_R (t)$.  Here the drive or
master system controls the response or slave system through the $x_D$ 
component.  The other component $x'_R$ is allowed to have different
initial conditions from that of  $x_R$.  The generalized schematic
set-up for this type of synchronization is shown in Fig. 24.

          One can easily see that if all the Lyapunov exponents of
the response system consisting of Eq. (55c) are less than zero,
then after the decay of the initial transients, $x'_R(t)$
will be exactly in step with $x_R(t)$.  The Lyapunov exponents for Eq. (55c) are called
the conditional Lyapunov exponents ($CLE$'s).  Also the response
system with negative conditional Lyapunov exponents is called a
stable subsystem or a stable response system.  If at least one
of the conditional Lyapunov exponents is positive, then
synchronization will not occur between the $x'_R$ and $x_R$ variables.  If
the $x'_R$ and $x_R$ subsystems are under perfect synchronization, then
the difference between the $x'_R$ and $x_R$ variables, $x_R^*  = x'_R - x_R$,
will tend to zero asymptotically, that is, $\lim_ {t\to\infty} x_R^* (t)
\rightarrow 0$.  Then we have
$$ \dot x _R^*(t) = \dot x'_R(t) - \dot x_R(t) = h(x_D, x'_R) - 
h(x_D, x_R). \eqno(56)$$
If the subsystem is linear, we have
     $$\dot x_R^*(t) = A x_R^*(t), \eqno(57)$$
where $A$ is a $(k \times k)$ constant matrix.  Let the eigenvalues of  $A$ 
be $(\lambda _1, \lambda_2, \cdots, \lambda_k)$.  The real parts of  
these eigenvalues are, by definition, the $CLE$'s we seek.

          If all of the $CLE$'s are negative, then $\lim_ {t\to\infty} x_R^*(t)=0$,
and the subsystems will synchronize; if there is a positive $CLE$, the 
subsystems will grow further apart as $t\rightarrow\infty$ and they will never 
synchronize.  Interestingly, an intermediate case occurs if one or 
more of the $CLE$'s are zero but none is positive; in this case, as 
$t\rightarrow\infty$, the trajectories of the subsystems will be separated by a 
fixed distance depending upon the initial conditions.

          If the response system is nonlinear in nature, then  the
conditional Lyapunov exponents are not so easily determined
and we must resort to numerical procedures to calculate them.  But
if the response system is linear (for example, a linear circuit with
passive elements), it is trivial to calculate the  $CLE$ in certain
cases.
          The above methodology has been successfully  applied  to
obtain chaos synchronization in many important nonlinear  systems
including the Lorenz system, the R\"{o}ssler system, the
hysteretic  circuit, Chua circuit, the driven Chua
circuit, the $DVP$ oscillator, phase-locked loops, and so on. For applications
to various systems and applications to secure communications we again
refer to [6].

\section {Conclusions}

In this set of lectures, we have tried to show how even seemingly
simple looking nonlinear oscillators can give rise to a very rich and
baffling variety of complex dynamics involving periodic and chaotic
motions.  Combined analytical, numerical and analog simulation studies
can help to unravel such complex dynamics in a systematic manner.
Recent developments also show clearly that chaotic dynamical behaviour
can be controlled algorithmically by minimal efforts so as to avoid
any harmful effects of chaos on the structure of the system.  Furthermore, 
identically evolving chaotic systems can be synchronized in a surprising
way through appropriate coupling and the concept can be profitably used
for spread spectrum communications.  Thus the study of even such simple
nonlinear systems can bring out an enormous amount of new knowledge on
dynamical systems.

Naturally, it will be of great interest to consider assemblies and
arrays of such oscillators with suitable couplings.  Such studies
can lead to interesting spatio-temporal patterns and chaos as well as to
new
structures of practical importance.  The associated mathematical 
analysis and physical realizations will be areas of intense future
investigations.

\section*{Acknowledgements}

I wish to thank Dr. K. Murali and Mr. P. Muruganandam for their help in the preparation
of this article.  This work forms part of a Department of Science and
Technology research project.
\newpage
\noindent {\bf Figure Captions}

\noindent {\bf Fig. 1(a)} (i) Solution curve $x(t)$ of Eq. (2) for 
$\omega_0^2$ = 1.0 (ii) Phase portrait in the $(x-\dot x)$ plane
corresponding to (i).

\noindent {\bf Fig. 1(b)} (i) Solution curve $x(t)$ of Eq. (5) for $\alpha = 0.5$,
$\omega_0^2$ = 1.0. (ii) Phase Portrait of (i).

\noindent {\bf Fig. 1(c)} (i) Solution curve $x(t)$ of Eq.(6) for $\alpha = 0.5$,
$f$ = 1.0, $\omega_0^2$ = 1.0, and $\omega = 1.0$. (ii) Phase portrait
of (i) (transients excluded).

\noindent {\bf Fig. 2} Resonance curve of the nonlinear oscillator (8) for
$\beta >0$.

\noindent {\bf Fig.3} Some typical attractors of a nonlinear oscillator.
(a) Trajectories ending at a stable focus. (b) Stable limit -
cycle attractor of period $T$ and angular frequency $\omega$ = $2\pi/T$.
(c) A quasiperiodic attractor of the Duffing-van der Pol oscillator.
(d) Chaotic attractor of the Murali-Lakshmanan-Chua circuit

\noindent {\bf Fig.4} A forced LCR circuit (8). Here $f(t) = F_s\sin
\omega_s t$.

\noindent {\bf Fig.5} (a) Wave form $v$ of circuit of Fig. 4.
(b) Trajectory
plot in the ($v-i_L)$ plane.  (c) Trajectory plot in the $(v-i_L)$ plane
of the experimental circuit of Fig. 4.

\noindent {\bf Fig.6} Analog simulation circuit of the driven Duffing
oscillator (double-well potential).

\noindent {\bf Fig.7} A typical three-segment piecewise-linear characteristic
curve of a nonlinear resistor $(N)$.

\noindent {\bf Fig.8} Circuit diagram of the simplest dissipative
nonautonomous circuit.

\noindent {\bf Fig.9} Various types of steady-state attractors of the Duffing
oscillator as a function of the forcing strength $f$.

\noindent {\bf Fig.10} Bifurcation diagrams of the Duffing oscillator
in the $(f-x)$ plane computed through numerical simulation.

\noindent {\bf Fig.11} (a): (i) Schematic diagram of a reverse bubble.
(b): (i) Schematic diagram of a bubble.  (a): (ii) Bifurcation diagram in the
range $0.43-0.48$ of the parameter $f$.
(b): (ii) Bifurcation diagram in the range $0.53$ $\leq f$ $\leq$ $0.62$

\noindent {\bf Fig.12} The $(f-\omega)$ phase diagram indicating regions of
different steady states exhibited by the double-well potential oscillator
for $\alpha$ = $0.1$, $\omega_0^2$ = $-0.5$, $\beta$ = $0.5$ in Eq.(25) with
$f(t) =  f \cos \omega t$.  (vertical lines): $LO$ symmetric;  (right
slanting lines): $LO$ unsymmetric;
shaded area: cross-well chaotic motion; cross-thatched $V$ areas: $SO$ occurs
outside $V$-shaped regions.  Adapted from Ref.[11].

\noindent {\bf Fig.13} Various types of steady state attractors
corresponding to Fig. 12. \\
\noindent {\bf Fig.14} Experimental results of the Duffing oscillator circuit
(Fig. 6). (i) Phase portrait in the $(x- \dot x)$ plane (ii) Lower trace:
solution curve $x(t)$, upper trace: forcing.

\noindent {\bf Fig.15} Bifurcation diagram of the BVP Eq.(17) in the
ranges (a) $A_1 \in  (0,1.8)$, (b) $A_1 \in  (0.6,0.8)$, and (c)
$A_1 \in (1.0,1.3)$.

\noindent {\bf Fig.16}  
The complete devil's staircase for Eq.(17) at $A_1 = 0.88$ with
$a = 0.8$, $b = 0.7$, $c = 0.09$, and $\omega= 1.0$

\noindent {\bf Fig.17} MLC circuit (28): (a) Numerical simulation results:
Phase portraits in $(x-y)$ plane:
(i) $f = 0.065$, Period-1 limit cycle; (ii) $f = 0.08$, Period-2 limit cycle;
(iii) $f = 0.091$, Period-4 limit cycle; (iv) $f = 0.1 $, One-band chaos;
(v) $f = 0.15$, Doubleband chaotic attractor.
(b) Experimental results:
Phase portrait $v$ versus $v_s(=R_si_L)$:
(i) $F = 0.0365V_{rms}$, Period-1 limit cycle; (ii) $F = 0.0549V_{rms}$,
Period-2 limit cycle; (iii) $F = 0.064V_{rms}$, Period-4 limit cycle;
(iv) $F = 0.0723V_{rms}$, One-band chaos; (v) $F = 0.0107V_{rms}$,
Doubleband chaotic attractor.

\noindent {\bf Fig.18} Codimension one bifurcations (see Table 2).

\noindent {\bf Fig.19} Codimension two bifurcations (see Table 3).

\noindent {\bf Fig.20} Regions of $LO$ attractor: computer simulation
and theoretical stability limits: $\alpha=0.1$ (Ref.[11]).

\noindent {\bf Fig.21} (a) Square lattice of poles found in typical
solution of free undamped Duffing oscillator $(\alpha = 0, f = 0)$.
(b) Singularity
distribution in the complex $t$-domain of a solution to the damped
oscillator.

\noindent {\bf Fig.22} Typical singularity distribution in complex $t$-domain
for a solution of the forced Duffing oscillator.
Accentuated poles (heavy dots) indicate commonly observed `tunnel'
structures (adapted from Ref.[13]).

\noindent {\bf Fig.23} Evolution of the variable $x$ of the $BVP$ oscillator.
(a) Limit cycle motion without control; (b) Attraction to the fixed
point $(x_s, y_s) = (-1.001, -0.376)$ under the control for $\epsilon = 0.1$.
Conversion of (c) chaotic motion to (d) a period -$T$ limit cycle and
(e) period-$2T$ limit cycle.  $A_{1f}$ is the value of $A_1$ associated
with the desired solution.

\noindent {\bf Fig.24} Schematic representation of cascading synchronization.
Drive and response $\#1$ (left slanting lines):   Variables used to drive the
response $\#1$. (right slanting lines):
The replica part of drive system used in response.  (blank): Drive variables
which are fed directly into the response  $\#1$.  Response  $\#2$.  Variables
which are fed directly into the response  $\#2$ from response  $\#1$. (left
slanting lines): The replica part of the drive system.

\newpage
\begin{table}
\caption{Summary of bifurcation phenomena of Eq. (25) with $\alpha=0.5$,
$\omega_0^2=-1.0$, $\beta=1.0$ and $\omega=1.0$.}
\begin{tabular}{rllc}
\hline
   &                     &                         &           \\
   & Value of $f$ & Nature of solution       & Attractor in the phase  \\
   &              &                          & space in the $(x, \dot x)$ plane \\
   &              &                          & (simulation results)  \\
   &              &                          & Numerical \hspace*{.5em} analog\\
   &              &                          & Fig. 9\hspace*{.2em} 
                                                \hspace*{2em} Fig. 14 \\
\hline
   &                 &                           &           \\
   &     $f=0$       & Damped oscillation to the &    (a)    \\
   &                 & stable focus at $x < 0$   &           \\
   & $0<f<0.34$      & Period-$T$ oscillation    &    (b)    \\
   & $0.34<f<0.358$  & Period-$2T$ oscillation   &    (c)    \\
   & $0.358<f<0.362$ & Period-$4T$ oscillation   &    (d)    \\
   & $0.362<f<0.37$  & Further period-doubling   &           \\
   & $0.37<f<0.4 $   & One-band chaos            &    (e)    \\
   & $0.4<f=0.42 $   & Double-band chaos         &    (f)    \\
   & $0.42<f $       & Chaos, windows, reverse   &  (g), (h) \\
   &                 & period-doublings, chaos,  &           \\
   &                 & boundary crisis, etc. (See &          \\
   &                 & bifurcation diagram of    &           \\
   &                 & Fig. 10)                  &           \\
   &                 &                           &           \\
\hline
\end{tabular}
\end{table}
\vskip 20pt
\begin{table}
\caption{Codimension one (one-parameter family of) bifurcations}
\begin{tabular}{|rl|c|c|}
\hline
   & Type of bifurcation & Model equation          & Nature of bifurcation \\
\hline
   &                     &                         &           \\
Flows  &                     &                         &           \\
1. & \underline{
     Saddle node}        & (i) $\dot x=\mu -x^3$   & Fig. 18a \\
   &                     & (ii) $\dot x=\mu +x^3$  & Fig. 18b \\
   &                     &  $x \in {\cal R}^1$,
                            $\mu \in {\cal R}^1$     &           \\
   &                     &                         &           \\
2. & \underline{
     Transcritical}      &     $\dot x=\mu -x^2$   & Fig. 18c \\
   &                     &                         &           \\
3. & \underline{
     Pitchfork}          &                         &           \\
   & (i) Supercritical   &    $\dot x=\mu x-x^3$   & Fig. 18d \\
   & (ii) Subcritical    &    $\dot x=\mu x+x^3$   & Fig. 18e \\
   &                     &                         &           \\
3. & \underline{
     Hopf}               &                         &           \\
   & (i) Supercritical   &    $\dot r=r(\mu-r^2)$  & Fig. 18f \\
   &                     &    $\dot \theta=1 $     &           \\
   &                     &    $r^2=x^2+y^2$        &           \\
   &                     &                         &           \\
   & (ii) Subcritical    &  $\dot r=r(\mu+r^2)$    & Fig. 18g \\
   &                     &   $ \dot \theta=1 $     &           \\
   &                     &                         &           \\
Maps   &                     &                         &           \\
4. & \underline{
     Flip}               &                         &           \\
   & (period doubling)   &$x_{n+1}=\mu x_n(1-x_n)$ & Fig. 18h \\
   &                     & $x\in [0,1)$,
                               $0\leq\mu\leq4$     &           \\
   &                     &                         &           \\
\hline
\end{tabular}
\end{table}
\vskip 20pt
\begin{table}
\caption{Codimension two (two-parameter family of) bifurcations}
\begin{tabular}{|rl|c|c|}
\hline
   & Type of bifurcation & Model equation          & Nature of bifurcation \\
\hline
   &                     &                         &           \\
1. & \underline{
     Cusp catastrophe}   & $\dot x=\lambda_1 +     
                           \lambda_2 x - x^3 $     & Fig. 19a  \\
   &                     & $x \in {\cal R}^1$,
                           $(\lambda_1, \lambda_2)
                           \in {\cal R}^2$         &           \\
   &                     &                         &           \\
2. & \underline{
     Degenerate Hopf}    & $\dot r=r(\lambda_1
                           +\lambda_2 r^2+\alpha r^4)$    & Fig. 19b  \\
   &                     & $\dot\theta = \omega
                            +{\bf O}(r^2)     $    &           \\
   &                     &                         &           \\
3. & \underline{
     Takens -Bogdanov }  & $\dot x=y$              & Fig. 19c  \\
   &                     & $\dot y=\lambda_1 x+
                         \lambda_2 y - x^3 - x^2y$ &           \\
   &                     &                         &           \\
\hline
\end{tabular}
\end{table}

\end{document}